\documentclass{pnastwoX}
\usepackage{graphicx}
\usepackage{amssymb,amsfonts,amsmath}

\contributor{Accepted for publication Proceedings
of the National Academy of Sciences of the United States of America}
\url{www.pnas.org/cgi/doi/10.1073/pnas.0709640104}
\copyrightyear{2008}
\issuedate{Issue Date}
\volume{Volume}
\issuenumber{Issue Number}

\begin{document}



\title{The Gemini Planet Imager: First Light}




\author{Bruce Macintosh     \affil{1}{Lawrence Livermore National Laboratory}
                            \affil{2}{Kavli Institute for Particle Astrophysics and Cosmology, Stanford University},
{James R. Graham}           \affil{3}{University of California, Berkeley},
{Patrick Ingraham}\affil{2}{},
{Quinn Konopacky}           \affil{4}{University of Toronto},
{Christian Marois}          \affil{5}{National Research Council of Canada},
{Marshall Perrin}           \affil{6}{Space Telescope Science Institute},
{Lisa Poyneer}\affil{1}{},
{Brian Bauman}\affil{1}{},
{Travis Barman}             \affil{21}{LPL, University of Arizona},
{Adam Burrows}              \affil{7}{Princeton University},
{Andrew Cardwell}           \affil{8}{Gemini Observatory},
{Jeffrey Chilcote}          \affil{9}{University of California, Los Angeles},
{Robert J. De Rosa}         \affil{10}{Arizona State University},
{Daren Dillon}              \affil{11}{University of California, Santa Cruz},
{Rene Doyon}                \affil{12}{Universit\'e de Montr\'eal and Observatoire du Mont-M\'egantic},
{Jennifer Dunn}\affil{5}{},
{Darren Erikson}\affil{5}{},
{Michael Fitzgerald}\affil{9}{},
{Donald Gavel}\affil{11}{},
{Stephen Goodsell}\affil{8}{},
{Markus Hartung}\affil{8}{},
{Pascale Hibon}\affil{8}{},
{Paul G. Kalas}\affil{3}{},
{James Larkin}\affil{9}{},
{Jerome Maire}\affil{4}{},
{Franck Marchis}            \affil{13}{SETI Institute},
{Mark Marley}\affil{22}{NASA Ames Research Center},
{James McBride}\affil{3}{},
{Max Millar-Blanchaer}\affil{4}{},
{Katie Morzinski}        \affil{14}{University of Arizona},
{Andew Norton}\affil{11}{}
{B. R. Oppenheimer}           \affil{15}{American Museum of Natural History},
{Dave Palmer}\affil{1}{},
{Jennifer Patience}\affil{10}{},
{Laurent Pueyo}\affil{6}{},
{Fredrik Rantakyro}\affil{8}{},
{Naru Sadakuni}\affil{8}{},
{Leslie Saddlemyer}\affil{5}{},
{Dmitry Savransky}          \affil{16}{Cornell University},
{Andrew Serio}\affil{8}{},
{Remi Soummer}\affil{6}{}
{Anand Sivaramakrishnan}\affil{6}{},\affil{15}{}
{Inseok Song}               \affil{17}{University of Georgia},
{Sandrine Thomas}           \affil{18}{NASA Ames},
{J. Kent Wallace}\affil{19}{Jet Propulsion Laboratory/California Institute of Technology},
{Sloane Wiktorowicz}\affil{11}{},
\and
{Schuyler Wolff}            \affil{20}{Johns Hopkins University}}

\contributor{Accepted from publication in the Proceedings of the National Academy of Sciences
of the United States of America}


\maketitle

\begin{article}

\begin{abstract}
The Gemini Planet Imager (GPI) is a dedicated facility for directly
imaging and spectroscopically characterizing extrasolar planets. It
combines a very high-order adaptive optics system, a
diffraction-suppressing coronagraph, and an integral field
spectrograph with low spectral resolution but high spatial
resolution. Every aspect of GPI has been tuned for maximum sensitivity
to faint planets near bright stars. During first light observations, we
achieved an estimated $H$ band Strehl ratio of 0.89 and a 5-$\sigma$
contrast of $10^6$ at 0.75 arcseconds and $10^5$ at 0.35
arcseconds. Observations of Beta Pictoris 
clearly detect the planet, Beta Pictoris b, 
in a single 60-second exposure with minimal
post-processing. 
Beta Pictoris b is observed at a separation of 
$434 \pm 6$~ milli-arcseconds (mas) 
and position angle $211.8 \pm
0.5^\circ$. Fitting the Keplerian orbit of Beta Pic b using the 
new position together with previous astrometry 
gives a factor of three improvement in most parameters over previous solutions. 
The planet orbits at a semi-major axis of $9.0^{+0.8}_{-0.4}$~AU
near the 3:2 resonance with the previously-known
6 AU asteroidal belt and is aligned with the inner warped disk. The
observations give a 4\% posterior probability of a transit of the planet in late
2017.
\end{abstract}

\section{Significance Statement}
Direct detection---spatially resolving the light of a planet from the
light of its parent star---is an important technique for characterizing
exoplanets.  It allows observations of giant exoplanets planets in locations like
those in our solar system, which are inaccessible to other methods. The Gemini
Planet Imager (GPI) is a new instrument for the Gemini South
telescope.  Designed and optimized only for high-contrast imaging, 
it incorporates advanced adaptive optics, diffraction
control, and a near-infrared spectrograph and an imaging polarimeter. 
During first light scientific
observations in November 2013, GPI achieved contrast performance that
is an order of magnitude better than
conventional adaptive optics imagers. 

\keywords{Adaptive Optics | extrasolar planets | astronomical instrumentation}

\dropcap{D}irect imaging is a powerful complement to indirect
exoplanet detection
techniques. In direct imaging, the 
planet is spatially
resolved from its star, allowing 
it to be independently
studied. This capability opens up new regions of parameter space, 
including sensitivity to planets at $>$ 5 AU. 
It also allows spectroscopic analysis of
the light emitted or reflected by the planet to determine its
composition\cite{Konopacky2013}\cite{Oppenheimer2013} 
and astrometry to determine the full Keplerian orbital 
elements\cite{Chauvin2012}\cite{Kalas2013}.

Imaging planets is 
extremely challenging---Jupiter is $10^9$
times fainter than our sun in reflected visible light. Younger
extrasolar planets are more favorable targets. During their formation,
planets are heated by the release of gravitational potential
energy. Depending on the exact formation process and initial
conditions, a 4 Jupiter-mass ($M_J$) planet at an age of 10 million
years could have a luminosity between $10^{-6}$ and $2 \times 10^{-5}
L_\odot$\cite{Marley2007}---but this is still a formidable contrast
ratio. To overcome this, astronomers have combined large telescopes (to
reduce the impact of diffraction), adaptive optics (to correct for
phase errors induced by atmospheric turbulence), and sophisticated
image processing
\cite{Lafreniere2007}\cite{Soummer2012}.
This recipe in various combinations had achieved
several notable successes
\cite{Kalas2008}\cite{Marois2008}\cite{Lagrange2010}\cite{Rameau2013}\cite{Kuzuhara2013}.
However, the rate of these discoveries remains
low\cite{Biller2013}\cite{Nielsen2013}\cite{Wahhaj2013}, in part
because the number of suitable young stars in the solar neighborhood
is low, and, for all but the closest stars, detection is limited
to $>20$ AU where planets may be relatively rare.  To move beyond
this limited sample, dedicated instruments are needed that are designed
specifically for high-contrast imaging. One such instrument is the
Gemini Planet Imager (GPI). GPI is a fully-optimized high
contrast AO facility deployed on an 8-m telescope and is almost
an order of magnitude more sensitive than the instruments of the
previous generation. 
With this powerful dedicated facility, large-scale surveys could
increase the sample of directly imaged giant planets to 25--50, or more
\cite{McBride2011}.
We present an overview of the design of GPI,
discuss its initial operation and performance, and show first-light
science results.

\section{Design of the Gemini Planet Imager}
At high dynamic ranges, the point spread function (PSF) of a bright
star imaged through a telescope is complex
\cite{Sivar2000}\cite{Perrin2003}, with light scattered by a variety
of different mechanisms and imperfections interacting coherently and
varying on different timescales. 
Atmospheric turbulence induces
large wavefront phase errors (and smaller intensity
fluctuations.) These are corrected (imperfectly) by the adaptive
optics (AO) system; the residual phase errors produce a halo of scattered
light. Instantaneously and monochromatically this halo is broken up
into individual speckles. Over long exposures, these speckles move
randomly and the atmospheric component of the halo reduces to a smooth
profile. However, if the measurements of the phase used to control the
adaptive optics are biased, the residual phase will have a non-zero
mean and the speckle pattern will have persistent components that
could be mistaken for a planet. These can be attenuated through PSF
subtraction \cite{Lafreniere2007}, but often the exact bias will vary
as, e.g., the mean seeing changes, resulting in a halo that changes on
timescales of minutes\cite{Hinkley2007}. An additional source of
persistent speckles are intensity variations 
either due to reflectivity variations or to propagation of wavefront
errors on surfaces not conjugate to the deformable mirrors (DM)
\cite{Marois2012}. 
Even in the absence of wavefront errors,
the finite aperture of the telescope causes diffraction 
that can swamp the planetary signal.

GPI is designed to deal with these errors through multiple
approaches. The AO system has a large number of degrees
of freedom and operates at high cadence to minimize residual
atmospheric turbulence. An apodized-pupil Lyot coronagraph (APLC)
\cite{Soummer2011} removes residual diffraction. Most importantly and
uniquely, the entire design of the optical system is intended to
minimize biases in the wavefront measurement and in turn minimize
quasi-static speckles.

The instrument was also designed to minimize the chromaticity of the
PSF---the APLC design sacrifices inner working angle for
achromaticity \cite{Soummer2011}, the number of transmissive elements
is kept to a minimum, individual mirrors are polished to $<1$~nm root mean square 
(RMS)
surface error\footnote{In the spatial frequency bandwidth of 4--24 cycles per pupil.}, 
and optics near focal planes are avoided to reduce
chromatic Fresnel effects. As a result, the residual speckle pattern
is highly correlated between different wavelength channels and can be
further attenuated through multi-wavelength
post-processing\cite{Marois2000}.  Table 1 summarizes the properties
of the instrument.

\subsection{Adaptive optics and optomechanical systems}
The adaptive optics system is intended to improve on previous
facilities in two respects---lower total wavefront error from dynamic
sources, 
and also lower quasi-static errors by an order of magnitude.
Typical current AO systems such as the 349-actuator system on the
Keck telescopes \cite{Wizinowich2000} has $\simeq$ 250~nm RMS of
dynamic wavefront error on a bright star \cite{vanDam2006}; GPI is
designed to achieve $\simeq 90$~nm. The static errors of current
systems are hard to estimate, but generally are 50 to 100~nm RMS.
For the Keck AO system this number is 113 nm
\cite{vanDam2004}; the design goal for GPI was 10~nm.

To achieve this goal, GPI adopts several novel features. The
high-order AO system has 43 subapertures across the 
the Gemini South pupil. A conventional piezoelectric DM with this
number of subapertures would have a diameter $\ge$ 20 cm, 
resulting in an instrument too large for the
Cassegrain focus. Instead, GPI uses a silicon MEMS DM
manufactured 
by Boston Micromachines corporation
\cite{Cornelissen2008}. This DM has 64 $\times$ 64 (4096) 
actuators---GPI uses a 48-actuator circle (including
slaved actuator rings) within this area. This compact (25.6-mm
diameter) device enables precision AO correction within a small
volume and mass (2200 kg). The MEMS DM has two noteworthy
limitations. First, the total displacement range is $\sim$ 4~$\mu$m,
insufficient to fully correct atmospheric turbulence. GPI therefore
also employs a piezoelectric, low-order DM, referred to as a ``woofer.'' 
Second, the 
MEMS DM has 
five defective actuators inside the controlled
pupil. Light from these bad actuators is blocked by the coronagraphic
Lyot mask.  Tip/tilt control is provided jointly by a piezoelectric
mount for the whole woofer DM ($<$ 30~Hz) and
actuation of the woofer surface itself ($>$ 30~Hz).

GPI employs a Shack-Hartmann wavefront sensor (AOWFS).
Such sensors are susceptible to aliasing of
errors outside the Nyquist range, which can
inject significant mid-frequency wavefront errors into the final
image. A spatial filter\cite{Poyneer2004} was
installed to mitigate this. The 
spatial filter is
adjustable with size set by the Nyquist criterion. 
The spatial filter can also be
used for calibration. By adjusting it to a pinhole 
we can inject a nearly perfect spherical wavefront into the AOWFS, allowing fast 
calibration of the dominant source of non-common-path aberrations in the system. 
GPI uses a computationally efficient Fourier transform reconstructor\cite{Poyneer2002} to
translate AOWFS slopes into DM commands. The control loop gain of each
Fourier mode is individually optimized
\cite{Poyneer2005} to minimize total wavefront error as atmospheric
turbulence and star brightness vary.

GPI operates at a Cassegrain focus and so is subject to a variable
gravity vector. 
The mechanical structure of GPI is therefore designed to minimize and actively
control flexure. Individual components such as the DM 
and lenslets remain optically registered to each other to within a few
microns.

\subsection{Coronagraph, science instrument, and IR wavefront sensor}
GPI uses an APLC, 
which suppresses diffraction by using a grey-scale mask to taper the transmission 
near the edges of the pupil. The transmission
profile of this mask is matched to the Fourier transform of the
hard-edged focal-plane stop (radius $2.8 \lambda / D$) such that
residual diffraction is channeled outside the telescope pupil at a
Lyot stop \cite{Soummer2011}. The masks were implemented with a
dithered half-tone pattern of 
10 $\mu$m 
metal dots on a glass
substrate; the focal-plane stop operates in reflection as a
gold-coated 
mirror with a small central hole. The machined
Lyot stops are located inside the cryogenic spectrograph. 
A grid of narrow, widely-spaced
lines printed onto the apodizer forms
a two-dimensional grating,
producing diffracted images of the central star in a square pattern.
These satellite images (Fig. 1) track the star's position and intensity,
facilitating astrometric and photometric calibration
data\cite{SivGrid2006}\cite{MarGrid2006}. A flexible and automated
data pipeline transforms raw data into calibrated
datacubes and starlight-subtracted images. Steps in this process
include removal of detector systematics, wavelength calibration, 
compensation for time-variable flexure,
assembly into multiwavelength datacubes, astrometric and photometric
calibration into physical units, and PSF calibration and
subtraction. This open-source software and extensive documentation
is freely available.

A near-infrared (IR) pointing and low-order wavefront sensor maintains the
star in the center of the focal plane mask hole
through feedback to the AO system. GPI also includes a high-accuracy
near-IR interferometer integrated with the coronagraph optics
to provide time-averaged measurements of the wavefront at the focal
plane stop, similar to that used with the
P1640 instrument\cite{Oppenheimer2012},
though this has not yet been tested on sky.

The science instrument is a near-IR integral field
spectrograph (IFS) \cite{Chilcote2012}. Similar to the OSIRIS and P1640
instruments, it samples each spatial location in the focal plane
using a lenslet array and disperses each resultant sample
using a prism producing $\sim 37,000$ individual
microspectra. Each spatial pixel corresponds to 0.0143 arc seconds on
the sky. In a single exposure, for each location in the field of
view the IFS produces a $\lambda/\delta\lambda\simeq$ 30--70
spectrum over one of the standard astronomical bands ($YJHK$, with $K$-band
split in two.)  
The IFS uses a Teledyne Hawaii 2RG detector.
Operating in up-the-ramp sampling mode it has a readout noise of 5
electrons RMS. It is cooled to $\sim$ 80 K with a pair of 15 W
Stirling-cycle cryocoolers.

\section{First-light results}
GPI was extensively characterized at the University of California, Santa
Cruz prior to shipping to Chile \cite{Macintosh2012}. In October 2013
it was attached to the bottom Cassegrain port of the Gemini South
telescope. After some brief alignment tests, first light occurred on
November 12, 2013 (UT); the first testing run continued until November 18, with a
second observing run from December 9 to 12.

\subsection{AO performance}
The instrument met its initial performance goals. The AO system locked
onto every target attempted, even in seeing with 
a Fried parameter of $r_0 \simeq 5$~cm.
The AO loop closed on stars $I\le 8$ mag. 
(GPI operates down to $I \simeq
10$ mag. in better seeing.) GPI can save extensive AO telemetry at the
full system frame rate, including DM commands, raw
wavefront sensor images and centroids, reconstructed phases, and
tip/tilt measurements and commands. These have been used to generate
an error budget (Table 2) for typical observations of $\beta$ Pictoris ($I=3.8$
mag.) from observations taken on December 11 (UT). The Fried parameter
$r_0$ was estimated to be 20~cm from measurements earlier in the night 
with a Differential Image Motion Monitor located outside the Gemini dome, 
corresponding to
better-than-average seeing. Most error terms are directly estimated
from AO telemetry, following methods used in \cite{vanDam2006}. 

Two error terms are worth further discussion. The first is a
quasi-static offset on the MEMS DM, consisting
primarily of a phase excursion in the outer two rings of
subapertures. Although this is the largest quasi-static term in the
error budget, it likely has only a small effect on final images, since
the APLC transmission for those subapertures is low. The likely cause
is a reconstruction of centroid patterns that do not correspond to a
physical wavefront (such as rotation of the whole centroid pattern), and
it will be mitigated through software changes. The second is a strong 60~Hz vibration. 
The Stirling cycle cryocoolers
on GPI induce vibration at harmonics of 60~Hz in the GPI mechanical
structure. Internal to GPI, these contribute about 6--8 mas
of RMS tip/tilt image motion at 120 and 180~Hz, as
measured with our artificial star unit. Observing actual stars through
the telescope, we see an additional 10 mas of 60~Hz image
motion, indicating a component external to GPI is vibrating. More
significantly, there is 100~nm (time-domain root-sum-squared of the spatial RMS) of
low-order wavefront vibration at 60~Hz, primarily focus, when the
cryocoolers are operating. This may indicate that the telescope's
mirrors, which are very lightweight, are being driven by the
GPI vibrations. We carried out measurements both with the cryocoolers
at full power and reduced to $\simeq 30 \%$ power. Modifications to
the cryocoolers in January 2014 are expected to significantly reduce
the vibration.

With the cryocoolers at full power, we estimate the RMS wavefront
error to be 134~nm; with the cryocoolers at minimum power, the RMS
wavefront error is 98~nm (exclusive of tip/tilt), corresponding to a
Strehl ratio of 0.87 at 1.65~$\mu$m. Although this is very good AO
performance, it is not vastly better than other high-order AO
systems. However, GPI significantly exceeds the raw image contrast of
such systems---we ascribe the difference to much better control of
quasi-static and systematic errors.

\section{First-light science: Beta Pictoris b}
Here we discuss observations of $\beta$ Pic. 
This young, nearby (19.4~pc), A6V star has a bright,
edge-on debris disk \cite{smith1984} and a directly-imaged
super-Jupiter, $\beta$ Pic b, orbiting at $\simeq$10 AU
\cite{Lagrange2010}.  Various asymmetries in disk structure have been
discovered \cite{lagage94a}\cite{kalas95a}, including a midplane warp
at $\sim$50 AU that has been attributed to a possible planetary
perturbation \cite{burrows95a}\cite{mouillet97a}\cite{heap00a}.

The planet has been detected by VLT/NACO\cite{Lagrange2010},
Gemini/NICI \cite{Boccaletti2013},
and the Magellan AO system \cite{Morzinski2014}. 
We observed the
planet in $H$ band (1.65~$\mu$m) on November 18 (2013) UT. We obtained
33 individual 60-second images in coronagraphic mode, with
the cryocoolers operated at minimum power. 
The planet was immediately
visible in a single raw 60-second exposure (Fig. 1.)
For comparison, a lower signal-to-noise $H$-band detection using NICI
\cite{Boccaletti2013} required 3962~s of exposure and
extensive PSF subtraction.

A set of 30 images were PSF-subtracted using a modified version of the
TLOCI algorithm \cite{Marois2014}. For each data cube wavelength
slice, the remaining wavelengths, and the images at the same
wavelength acquired at a different time, are used as PSF reference images.
A simulated planet having either a flat spectrum similar to
L-dwarfs or a CH$_4$-dominated spectrum similar to 
T-dwarfs is used to exclude reference images with more than
10\% self-overlap of the planet with the image being processed.  The
selected reference images are then used to subtract the stellar noise
by performing a LOCI least-squares fit constrained to positive
coefficients. All star-subtracted data cubes are then rotated to put
North up and median combined (Fig. 2.)

We evaluate the contrast of the final image by calculating the
standard deviation of the intensity in concentric annuli about the
star. Figure 3 shows a plot of contrast for a 30-minute set of GPI
data. In 30-minute exposures, GPI is performing $\sim 2$ magnitudes
better than typical contrast obtained at the Keck Observatory. 

\subsection{Astrometry} 
We use these images to constrain the orbit of $\beta$ Pic b. 
Most astrometric calibrators are unsuitable for GPI
because of the small field of view (2.8 arc seconds on a side)
and also because for low-contrast ratio binary stars
the secondary will either saturate the science detector 
or light from the secondary will perturb the wavefront sensor. 
Therefore, we calibrated the scale and orientation of GPI
using other extrasolar planets as references. We compared the 
offset between HR 8799c and HR 8799d measured 
at the W.M. Keck Observatory in October 2013 with
GPI observations of this system.
The position of $\beta$ Pic b is measured from the combined TLOCI
image. The position of $\beta$ Pic b 
is estimated in each raw 60-s image
from the center of the square pattern defined by 
the four reference-grid spots. The measured separation of $\beta$ Pic b is $434
\pm 6$ mas at a position angle of $211.8 \pm 0.5$ degrees.

We fit the new GPI astrometric measurements presented here together
with the compilation of data in \cite{Chauvin2012}. We use 
Markov Chain techniques described in \cite{Kalas2013}.
The adaptive
Metropolis-Hasting sampler used in \cite{Kalas2013} has been replaced
by an affine invariant sampler\cite{goodman2010ensemble}\cite{Foreman2013}. 
Figure 4 shows an ensemble of trajectories computed using orbital
elements sampled from the joint posterior distribution.  The results
show a good fit with a posterior expectation value of $\chi^2_\nu =
0.76$ ($\nu = 14$).

The GPI data point implies that $\beta$ Pic b has recently
turned around on its orbit and roughly one half an orbital period has
elapsed since the first astrometric measurement in 2003. 
The combination of small measurement errors and the lever arm provided by
extended orbital coverage yields the best estimate 
to date of the semi-major
axis. Our new measurement places $\beta$ Pic b at
a semi-major axis of $9.0_{-0.4}^{+0.8}$~AU, 
with a corresponding period of $20.5_{-1.4}^{+2.9}$~yr.  
Beta Pic b will remain observable ($>$
0.2 arcseconds) at least until March 2016 and will likely reappear before
October 2019.  A possible transit of the planet across the star was
observed in 1981\cite{Lecavelier1995}. Our estimated orbital
inclination is 
$90.7\pm0.7^\circ$; transits of $\beta$ Pic will
occur if the inclination is within 0.05$^\circ$ of edge on.  
The next opportunity to observe such a
transit by $\beta$ Pic b is in September through
December 2017 (68\% confidence).

The orbit has a small but nonzero eccentricity ($e =0.06_{-0.04}^{+0.07}$).  
The corresponding range between periaspe and apoapse
(8.8--9.4~AU; see Table \ref{tab-percentiles}) falls neatly 
between the 6.4 AU and 16 AU asteroidal 
belts\cite{Weinberger2003}\cite{Wahhaj2003}\cite{oka2004}
and places the inner belt close to the 3:2 commensurability with the
orbital period of $\beta$ Pic b 

Our results agree
with \cite{Fitz2009}\cite{Chauvin2012}---the chief difference is the 
improvement in the confidence
intervals (see Table \ref{tab-percentiles}).
Distinct peaks are beginning to emerge in the posterior distribution
for epoch of periapse and for $\omega$ (which is significant for
identifying the perturber of the in-falling evaporating bodies.)
The 68\% confidence interval for $i$ drops from $3. 4^\circ$ to $1. 36^\circ $. 
The posterior probability of transit has increased by 
a factor of 2.2 relative to \cite{Chauvin2012}, but the
chance that $\beta$ Pic b transits the star
is still relatively small (4\%).

The longitude of the ascending node now has a confidence interval of
$0.^\circ 9$ compared to $3^\circ$ previously. The corresponding PA
is $211.6 \pm 0.^\circ 5 $ and clearly misaligned with the main disk
(PA = $209 \pm 0.^\circ 3$) but more nearly aligned with the inner
warped component\cite{Chauvin2012}\cite{Lagrange2012}.

These results show the power of a dedicated high-contrast imager; GPI
achieves a given 
contrast ratio 
sensitivity $\sim50$ times faster than the best
previous-generation systems. A 600-star survey of young nearby stars
will begin in 2014, with the goal of producing a sample of directly
imaged planets that: (a) spans a broad range of temperatures, ages, and
masses, opening up new areas of planetary atmosphere phase space; and (b)
probes the range of semi-major-axes and stellar ages inaccessible to
Doppler and transit surveys to produce statistical constraints on
planet-formation model. Simulations predict that 
this survey will discover
25-50 exoplanets \cite{McBride2011} with masses as low as that of
Jupiter and at separations as close as 3-5 AU. Using a polarimetry
mode, the survey will also map debris disks. GPI will
be available to the astronomical community for guest observer
projects, ranging from studying young solar system objects to the
outflows from evolved stars. With the advent of GPI and similar
dedicated facilities, high contrast imaging, spectroscopy and
polarimetry will open up a new segment of other solar systems to
characterization.

\begin{acknowledgments}
We acknowledge financial support of the Gemini Observatory, 
the NSF Center for Adaptive Optics at UC Santa Cruz, 
the NSF (AST-0909188; AST-1211562), NASA Origins (NNX11AD21G; NNX10AH31G), 
the University of California Office of the President (LFRP-118057), and
the Dunlap Institute, University of Toronto. Portions of this work were 
performed under the auspices of the U.S. Department of Energy by Lawrence 
Livermore National Laboratory under Contract DE-AC52-07NA27344
and under contract with the 
California Institute of Technology/Jet Propulsion Laboratory 
funded by NASA through the Sagan Fellowship Program executed by the 
NASA Exoplanet Science Institute.
We are indebted to the international team of engineers and scientists 
who worked to make GPI a reality. We would especially like to recognize the unique 
contributions of Gary Sommargren, Steven Varlese, Christopher Lockwood, 
Russell Makidon, Murray Fletcher, and 
Vincent Fesquet, who passed away during the course of this project.

\end{acknowledgments}

\end{article}




\begin{table}[h]
\caption{Properties of the Gemini Planet Imager}
\begin{tabular}{lr}
Property & Value \\
\hline
Deformable mirror   & $64\times64$ MEMS\tablenote{1493 active actuator}   \\
AO Wavefront sensor & Spatially-filtered Shack-Hartmann\tablenote{1 pixel guard bands}  \\
AOWFS format        & $N=43$ subaps $2\times2$ pixels \\
AO limiting mag.    & $I=10$ mag \\
Coronagraph IWA     & $2.8\lambda/D$ \tablenote{0.112 arcsec at $1.65\ \mu$m}\\
IFS wavelength range    & 1.0--2.5 $\mu$m\tablenote{Split into five bands}   \\
IFS resolving power\tablenote{Also supports polarimetry} & $\lambda/\delta\lambda=30-70$   \\
IFS spatial sampling             & 0.0143 arcseconds per pixel  \\
IFS field of view                & 2.78 $\times$ 2.78 arcseconds  \\
\hline
\end{tabular}
\end{table}

\begin{table}[h]
\caption{Adaptive optics error budget for observations of Beta Pictoris}
\begin{tabular}{lr}
Error term & RMS value (nm)  \\
\hline
Fitting error  & 60\tablenote{Estimated from $r_0$ =18~cm} \\
Servo-lag error & 50 \tablenote{Calculated from AOWFS residuals} \\
AOWFS measurement noise & 6$^\dagger$ \\
Static edge excursions & 31 \tablenote{Calculated from average DM shape} \\
Other static residuals & 13$^\dagger$\\
60 Hz vibration & 100  \tablenote{CCRs at full power} \\
60 Hz vibration & 30--50\tablenote{CCRs at 30\% power} \\
Residual non-common-path & 30 \tablenote{Estimated from internal calibration source, mostly low frequency}\\
\hline
Total & 134 $^{\S}$ \\
Total & 98 $^{\P}$ \\
\hline
\end{tabular}
\end{table}

\begin{table}
\caption{Keplerian orbital elements.  The quoted value is the
 median (50th percentile) of the marginalized posterior distributions
 with errors representing the 68\% confidence interval (16 and 84th
 percentiles.) Derived quantities---period, $P$,
 periapse and apoapse distances are also listed.}
\begin{tabular}{@{\vrule height 10.5pt depth4pt  width0pt}lrl}
Orbital Element & Value & Units\\
\hline
$a$       &    $9.04_{-0.41}^{+0.82 }$  & [AU]   \\
$\Omega$  &    $-148.36\pm 0.45$        & [deg.] \\
$i$       &    $90.69_\pm 0.68$         & [deg.]  \\
$e$       &    $0.06_{-0.04 }^{+0.07 }$ &        \\
$P$       &    $20.5_{-1.4 }^{+2.9 }$   &[yr]    \\
$(1-e)a$  &    $8.75_{-0.81 }^{+0.24 }$ &[AU]    \\
$(1+e)a$  &    $9.39_{-0.33 }^{+1.55 }$ &[AU]    \\
\hline
\label{tab-percentiles}
\end{tabular}
\end{table}

\begin{figure}
\label{H-60-sec}
\includegraphics[width=4in]{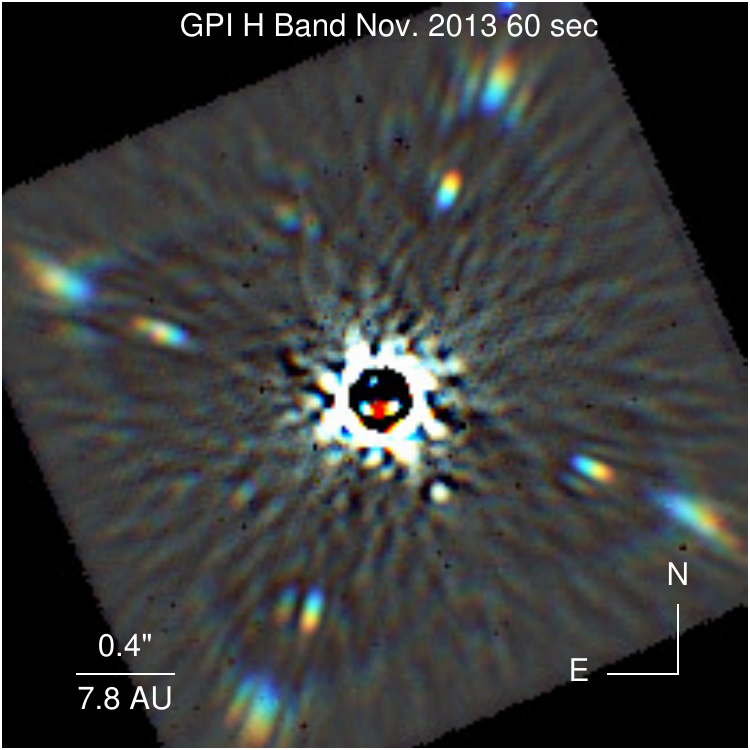}
\caption{RGB color composite of a single 60-second $H$ band (1.5--1.8~$\mu$m) 
  GPI image of Beta Pictoris. The short, medium and long
  segments of $H$ band are mapped to RGB. The image has been high-pass
  filtered to remove diffuse background light but no PSF subtraction
  has been applied. The four sharp spots at 1:00, 4:00, 7:00 and 10:00 o'clock
  are diffracted images of the star generated by a reference grid
  inside GPI.}
\end{figure}

\begin{figure}
\label{H-30-min}
\includegraphics[width=4in]{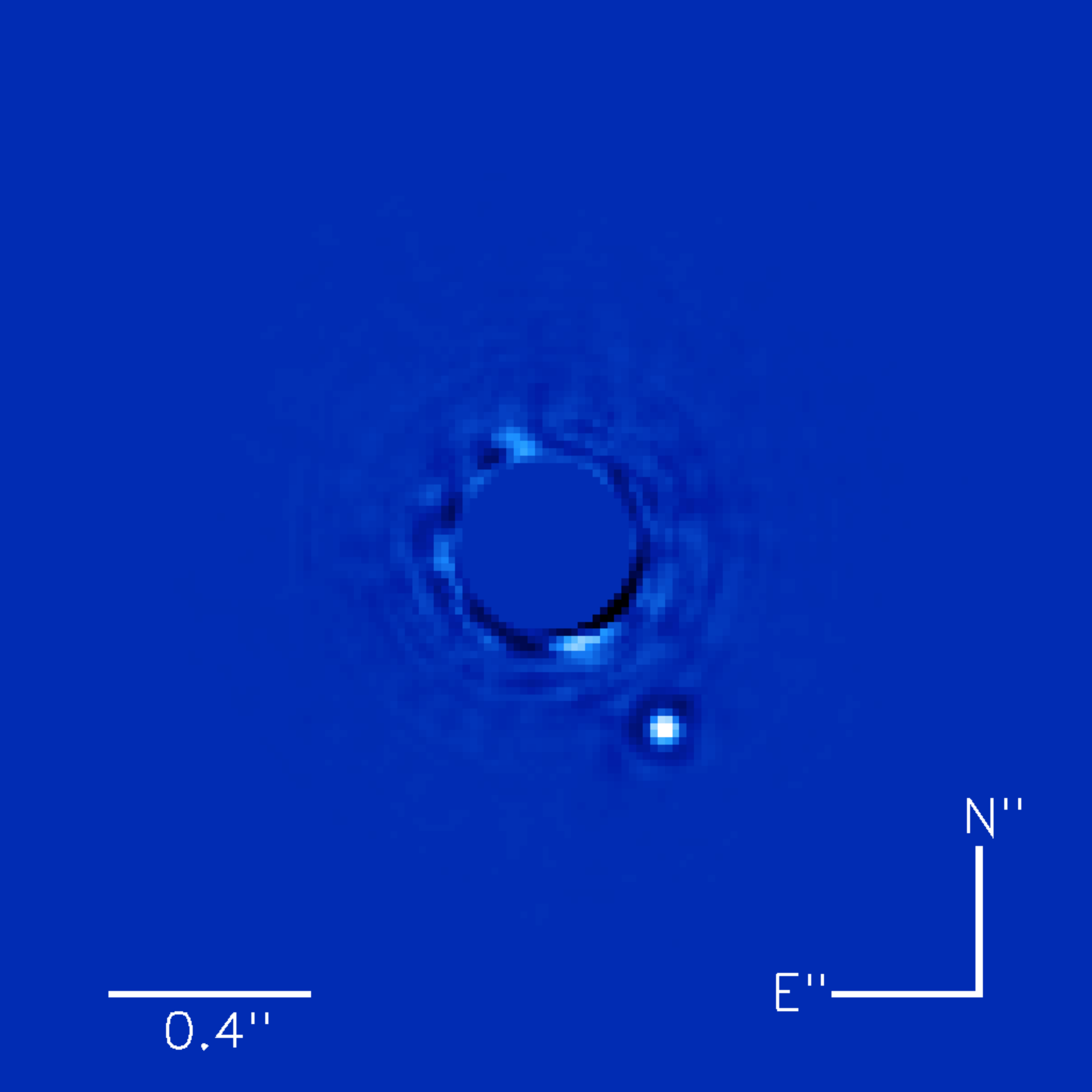}
\caption{Combined 30-minute GPI image of Beta Pictoris. The spectral
  data has been median-collapsed into a synthetic broadband 1.5--1.8~$\mu$m 
  channel. The image has been PSF subtracted using angular and
  spectral differential techniques. Beta Pictoris b is detected at a
  signal-to-noise of $\sim100$}
\end{figure}

\begin{figure}
\label{contrastvsep}
\includegraphics[width=4in]{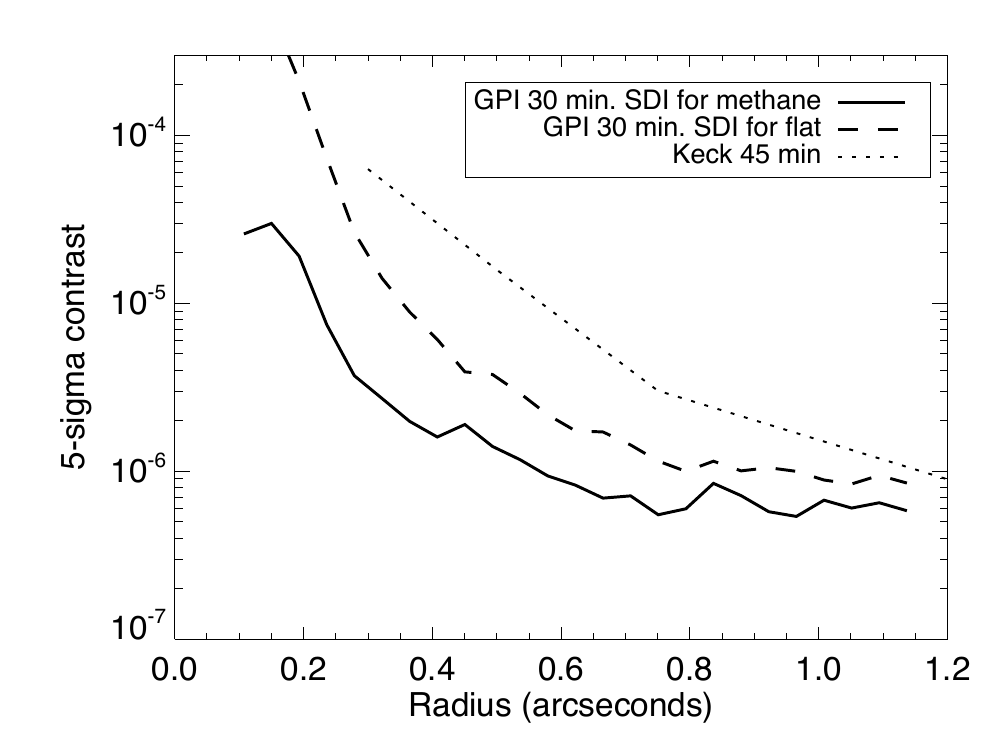}
\caption{Contrast vs. angular separation at $H$ (1.6 $\mu$m) for a PSF-subtracted
  30-minute GPI exposure. Contrast is shown for PSF subtraction based on either a flat spectrum similar to a L dwarf or a methane-dominated spectrum (which allows more effective multi-wavelength PSF subtraction.) For comparison, a 45-minute 2.1~$\mu$m Keck
  sequence is also shown. (Other high-contrast AO imaging setups such as Subaru HiCIAO, Gemini NICI, and VLT NACO have similar performance to Keck.) }
\end{figure}

\begin{figure}
\label{fig-data-fit}
\includegraphics[width=4in]{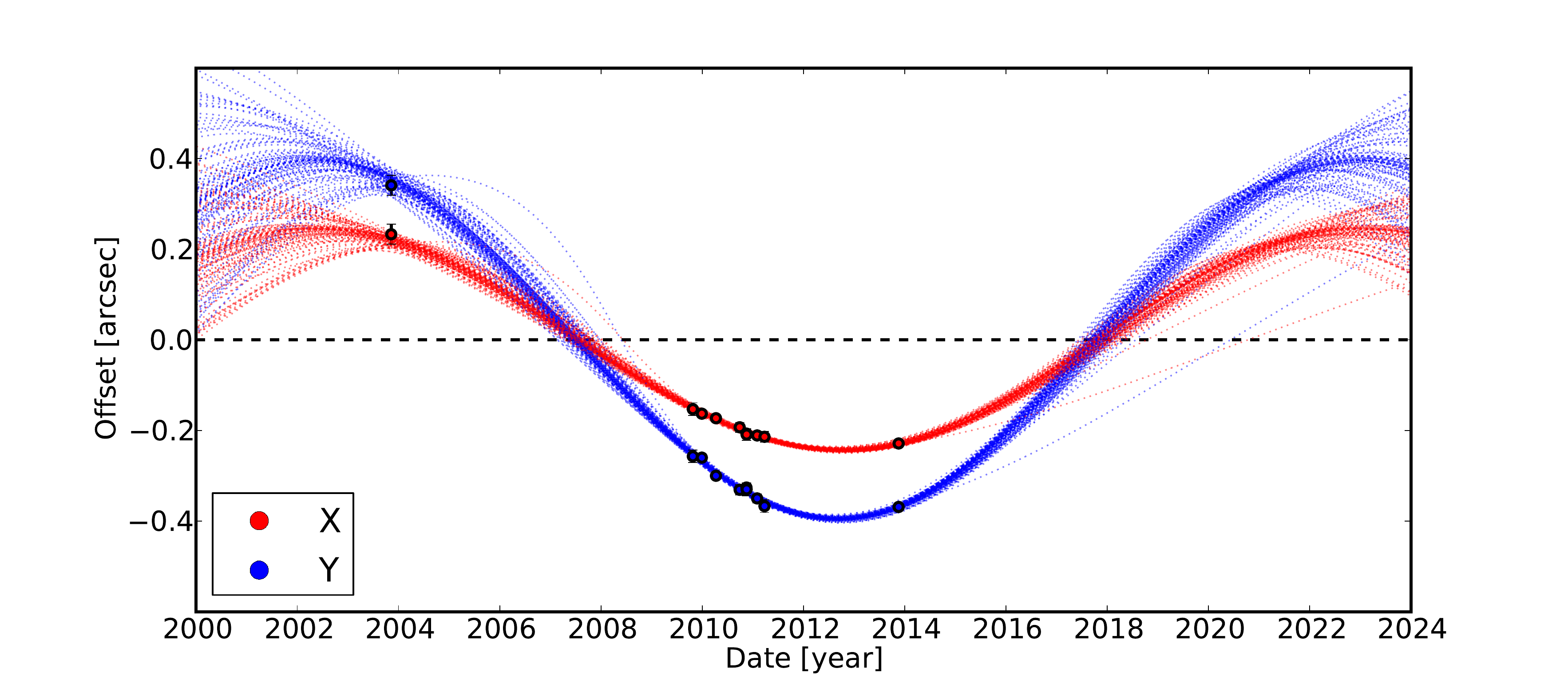}
\caption{The projected separation of $\beta$ Pic b relative to $\beta$ Pic in
celestial coordinates. Data prior to the GPI point in late 2013 are
from\cite{Chauvin2012}. The GPI measurement shows that $\beta$ Pic b
has passed quadrature and allows a prediction of conjunction in
Sept-Dec 2017. The red and blue trajectories show 100 solutions drawn
from the posterior distribution of likely orbital solutions determined
from our Monte Carlo Markov chain analysis of these data.}
\end{figure}

\begin{figure}
\label{avse-covariance}
\includegraphics[width=4in]{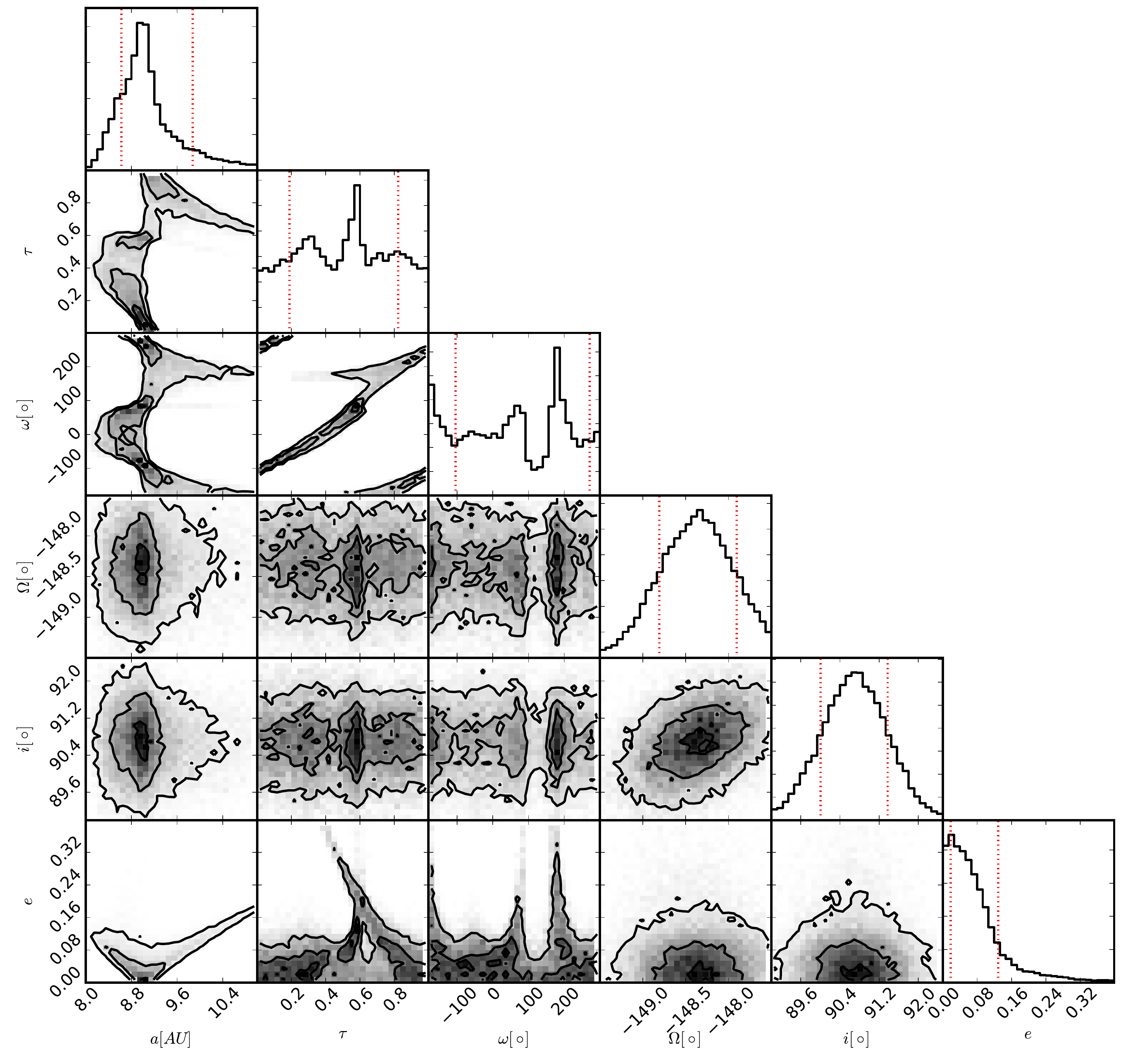}
\caption{Posterior distribution of the orbital elements
of $\beta$ Pic b: semi-major axis, $a$,
epoch of periapse, $\tau$ (in units of the orbital period),
argument of periapse, $\omega$, 
argument of the ascending node, $\Omega$,
inclination, $i$, and eccentricity, $e$. 
The plot shows the joint distributions 
as contours (0.1, 0.5, and 0.9) and marginalized
probability density functions as histograms. 
The well represented degeneracy in $\omega$, e.g., see
$\omega$ versus $a$,  
is good evidence of reliable sampling of the 
posterior distribution. 
Vertical dotted lines in the histograms denote the 68\% confidence interval.}
\end{figure}

\end{document}